\documentclass[useAMS,usenatbib]{mn2e}
\usepackage{epsfig}
\usepackage{amsmath}

\newcommand{\be}{\begin{equation}}
\newcommand{\beq}{\begin{equation}}
\newcommand{\ba}{\begin{eqnarray}}
\newcommand{\ee}{\end{equation}}
\newcommand{\eeq}{\end{equation}}
\newcommand{\ea}{\end{eqnarray}}

\newcommand{\apj}{ApJ}
\newcommand{\apjl}{ApJL}
\newcommand{\mnras}{MNRAS}
\newcommand{\aj}{AJ}

\newcommand{\nat}{{\it Nature}}
\newcommand{\pasj}{PASJ}
\newcommand{\kms}{{\rm ~km\,s^{-1}}}
\newcommand{\GC}{{\rm GC}}
\newcommand{\LG}{{\rm LG}}
\newcommand{\ZoA}{{\rm ZoA}}
\newcommand{\CMB}{{\rm CMB}}

\def\lsim{~\rlap{$<$}{\lower 1.0ex\hbox{$\sim$}}}

\def\gsim{~\rlap{$>$}{\lower 1.0ex\hbox{$\sim$}}}

\title[Local Group Dynamics]{Dynamical Constraints on the Local Group
from the CMB and 2MRS Dipoles}

\author[Loeb \& Narayan]{Abraham Loeb and Ramesh Narayan\\ Astronomy
Department, Harvard University, 60 Garden St., Cambridge, MA 02138\\Email:
aloeb@cfa.harvard.edu, rnarayan@cfa.harvard.edu}

\begin{document}


\maketitle

\label{firstpage}
\begin{abstract}

We place constraints on the dynamics of the Local Group (LG) by
comparing the dipole of the Cosmic Microwave Background (CMB) with the
peculiar velocity induced by the 2MRS galaxy sample. The analysis is
limited by the lack of surveyed galaxies behind the Zone of Avoidance
(ZoA). We therefore allow for a component of the LG velocity due to
unknown mass concentrations behind the ZoA, as well as for an unknown
transverse velocity of the Milky Way relative to the Andromeda galaxy.
We infer extra motion along the direction of the Galactic center
(where Galactic confusion and dust obscuration peaks) at the $95\%$
significance level.  With a future survey of the ZoA it might be
possible to constrain the transverse velocity of the Milky Way
relative to Andromeda.

\end{abstract}


\section{Introduction}

The amplitude of the dipole of the cosmic microwave background (CMB) is two
orders of magnitude larger than the characteristic amplitude of the
higher-order multipoles of its anisotropies \citep{Kogut}.  It is therefore
widely believed that the CMB dipole originates from the Doppler effect of
our peculiar velocity, which is induced by inhomogeneities in the local
universe \citep{Huchra,Conklin,Henry} rather than by a primordial origin
\citep{Gunn,Piran}. Indeed, 21cm surveys employing the Tully-Fisher
relation for distance calibration have inferred that the peculiar velocity
of the Local Group (LG) relative to distant galaxies converges within a
distance of $\sim 5,000~{\rm km~s^{-1}}$ or $\sim 70$Mpc
\citep{Giovanelli,DG}
\footnote{We note that the distance at which dipole
convergence is achieved is still controversial. Surveys of galaxy clusters
imply large convergence distances (Plionis et al 2000; Basilakos \& Plionis
2006; Ebeling et al. 2002, 2005; Kocevski 2005) but are affected by the
strong bias and Poisson fluctuations of clusters relative to the underlying
matter distribution. In this paper we assume that convergence is
reached within the maximum 2MRS distance of 280Mpc or $\sim 20,000
\kms$ (see \S 4.2 for the justification of this assumption).}.  This
important result confirms the notion that the peculiar velocity is induced
within that distance, since otherwise the distant galaxies would also be
moving relative to the CMB together with the LG.

Surveys of galaxies in the local universe have attempted over the past two
decades to explain the amplitude and direction of the CMB dipole within a
distance of $\gsim 100$Mpc \citep{LB,Strauss,conf1,conf2,conf3,conf4} The
adopted method assumes that: {\it (i)} the LG peculiar motion is induced by
gravity; and {\it (ii)} the amplitude of inhomogeneities in the
distribution of the observed light from galaxies traces the underlying mass
distribution on large spatial scales with a constant bias factor $b$. The
latest results, based on the 2 Micron All-Sky Redshift Survey (2MRS,
Erdo{\u g}du et al. 2006), show convergence of the flux-weighted dipole in
the galaxy survey out to $\sim 150~{\rm Mpc}$ but still indicate a
discrepancy of $24^\circ \pm 8^\circ$ with the direction of the CMB dipole.

The main limitation of the 2MRS sample results from the lack of sample
galaxies behind the Zone of Avoidance (ZoA), a strip around the Milky Way
disk where confusion and dust obscuration compromise the survey
efficiency. For lack of better information, the 2MRS analysis is also based
on the assumption that the Milky-Way galaxy is moving radially towards the
Andromeda galaxy (M31) with no transverse motion \citep{Vanden}. Our goal
in this paper is to constrain the unknown peculiar velocity of the
LG within the ZoA as well the unknown transverse speed of the Milky-Way
relative to Andromeda, by requiring a match between the 2MRS and CMB
dipoles.  The contribution of mass concentrations outside the survey volume
of 2MRS can be ignored based on the success of Tully-Fisher surveys in
converging to the CMB dipole within the same volume \citep{Giovanelli,DG}.

The outline of this paper is as follows. We first summarize the
existing data on the LG velocity from the CMB and Galactic measurements (\S
2) as well as from the 2MRS analysis (\S 3). We then compare the results
from the CMB and 2MRS data sets and interpret our results in the context of
transverse motion of the Milky Way relative to the LG (\S 4.1), structure
beyond the extent of 2MRS (\S 4.2), and
nearby structure within the ZoA (\S 4.3).  We conclude that the last
effect is the most likely explanation for the discrepancy between the CMB
and 2MRS dipoles.  We also derive the
likelihood function for the bias parameter $b$ of the 2MRS galaxies.
Finally, we discuss the implications of our results in \S 5. Throughout our
analysis, we use Galactic Cartesian coordinates in which the $x$-axis is
oriented towards the Galactic Center, i.e., towards longitude $l=0$ and
latitude $b=0$, $y$ is in the direction $l=90^\circ,b=0$, and $z$ is in the
direction $b=90^\circ$.

\section{CMB Dipole and Galactic Measurements}

The velocity of the Sun with respect to the CMB is $369.5\pm3.0 \kms$
towards $l=264.4^\circ\pm0.3^\circ$, $b=48.4^\circ\pm0.5^\circ$ (Kogut
et al. 1993).  In the Cartesian Galactic coordinate system,
\begin{equation}
\vec{v}_{\odot-\CMB} = (-23.9\pm1.3,-244.1\pm3.1,276.3\pm3.1) \kms.
\label{vSunCMB}
\end{equation}
Here and elsewhere, we add errors in quadrature and ignore possible
correlations between errors.

Courteau \& van den Bergh (1999) estimate the velocity of the Sun with
respect to the center of mass of the Local Group (LG) to be $306\pm18
\kms$ towards $l=99^\circ\pm5^\circ, b=-4^\circ\pm4^\circ$:
\begin{equation}
\vec{v}_{\odot-\LG} = (-47.8\pm26.5,301.5\pm18.3,-21.3\pm21.3) \kms.
\label{CvdB}
\end{equation}
The model used for this derivation assumed statistical isotropy of the
velocity distribution of the LG galaxies, which may not be satisfied
since most LG members are low-mass galaxies concentrated around the
Milky Way or Andromeda.  Therefore, instead of using this estimate we
will sum the best estimates for the velocity of the Sun with respect
to the Galactic Center (GC), $\vec{v}_{\odot-\GC}$, and the velocity
of the GC with respect to the LG, $\vec{v}_{\GC-\LG}$.

Reid \& Brunthaler (2004) have measured the proper motion of Sgr A$^*$
to be $-6.379\pm0.026 ~{\rm mas\,yr^{-1}}$ in longitude and
$-0.202\pm0.019 ~{\rm mas\,yr^{-1}}$ in latitude.  Since Sgr A$^*$ is
almost certainly at rest with respect to the GC, its proper motion is
entirely due to the component of the Sun's motion in the $y$ and $z$
directions.  To get these velocity components it is necessary to
specify the distance to the GC, for which we adopt the estimate of
Eisenhauer et al. (2003): $R_0 = 7.94\pm0.42 ~{\rm kpc}$.  For the $x$
component of the Sun's velocity we take the estimate of Dehnen \&
Binney (1999): $10.0\pm0.36 \kms$.  Thus,
\begin{equation}
\vec{v}_{\odot-\GC} = (10.0\pm0.36,240.1\pm12.7,7.6\pm0.8) \kms.
\label{vSunGC}
\end{equation}

The radial velocity of the Sun towards M31 is $-297 \kms$ according to
Mateo (1998), which is slightly different from the value given by
Courteau \& van den Bergh (1999), namely $301 \kms$.  We 
adopt Mateo's value and include a generous error estimate: $-297\pm5
\kms$.  The direction to M31 is $l=121.2^\circ, b=-21.6^\circ$. The
unit vector in this direction is
\begin{equation}
\hat{n}_{{\rm M31}} = (-0.4816,0.7953,-0.3681).
\end{equation}
The component of the Sun-GC velocity parallel to this unit vector is
$183.3\pm10.1 \kms$.  The remainder of the line-of-sight velocity
between the Sun and M31, $297\pm5\kms$, must be due to the relative
motion between the GC and M31.  Thus, we find
\begin{equation}
v_{||,\GC-{\rm M31}} \equiv \vec{v}_{\GC-{\rm M31}}\cdot\hat{n}_{{\rm M31}}
=113.7\pm11.3 \kms.
\end{equation}
Various estimates of the total mass of M31 place it between 4/3 (Mateo
1998) and 3/2 (Courteau \& van den Bergh 1999) of the mass of the
Milky Way.  Thus, we estimate the parallel component of the Galaxy's
velocity with respect to the center of mass of the LG to be
\begin{equation}
v_{||,\GC-\LG} = 66.6 \pm 6.7 \kms.
\label{vparGCLG}
\end{equation}

Combining equations (\ref{vSunGC}) and (\ref{vparGCLG}), and assuming
that the Milky Way has no transverse velocity with respect to the LG,
we calculate the velocity of the Sun with respect to the LG to be
\begin{equation}
\vec{v}_{\odot-\LG}=(-22.1\pm3.2,293.1\pm13.8,-16.9\pm2.6) 
~\kms.
\end{equation}
We note that this estimate agrees with that given by Courteau \& van
den Bergh (1999) in equation (\ref{CvdB}), to within the errors.  

Combining our estimate of $\vec{v}_{\odot-\LG}$ with the measured
velocity of the Sun with respect to the CMB, $\vec{v}_{\odot-\CMB}$
(eq. \ref{vSunCMB}), we obtain
\begin{equation}
\vec{v}_{\LG-\CMB}=(-1.8\pm3.5,-537.2\pm14.1,293.2\pm4.0)
~\kms . 
\label{vLGCMB1}
\end{equation}

\section{Local Group Motion from 2MRS}

The 2 Micron All-Sky Redshift Survey (2MRS) includes a sample of
infrared-selected galaxies out to an expansion velocity of $\sim
20,000 \kms$.  By assuming a constant mass-to-light ratio per
unit volume, the light distribution of these galaxies has been used to
derive the gravitational acceleration of the LG due to structure in
the local universe (Erdo{\u g}du et al. 2006).  From the {\it
flux-weighted} results in the {\it CMB frame} reported by Erdo{\u g}du
et al. (2006) in their Table 1, the expected velocity of the LG with
respect to the CMB is $(1620\pm327)f(\Omega_m)/b \kms$ towards the
direction $l=247^\circ\pm11^\circ, b=37^\circ\pm10^\circ$.  Here,
$f(\Omega_m) \approx \Omega_m^{0.6}$, where $\Omega_m$ is the matter
density of the universe, and $b$ is the mean bias factor of the
galaxies contributing to the acceleration of the LG.

Tegmark et al. (2006) have combined WMAP and SDSS data to estimate
$\Omega_m = 0.24 \pm 0.02$, which gives $\Omega_m^{0.6} =
0.425\pm0.021$.  The 2MRS data then yield the following prediction for
the velocity of the LG with respect to the CMB,
\begin{equation}
b\vec{v}_{\LG-\CMB}=(-214.8\pm110.6, -506.1
\pm131.1,414.4\pm128.9) \kms . 
\label{vLGCMB2}
\end{equation}
The error estimates include shot noise but not the effect of the
missing information in the ZoA.  The latter is discussed in \S~4.3.

A comparison of the velocity estimates given in (\ref{vLGCMB1}) and
(\ref{vLGCMB2}) suggests that, regardless of the value of the bias
factor $b$, there is a substantial discrepancy.  Let us adjust $b$ so
as to minimize the magnitude of the discrepancy.  The minimum occurs
at $b=1.056$.  The corresponding velocity discrepancy between
(\ref{vLGCMB1}) and (\ref{vLGCMB2}) is then
\begin{equation}
\Delta\vec{v} = (201.6\pm104.8, -57.9
\pm124.9,-99.2\pm122.1) \kms .
\label{vLGCMB3}
\end{equation}
The deviation is most significant in the $x$-components of the two
velocities.  Erdo{\u g}du et al. (2006), who noted this discrepancy,
offered a number of possible
explanations for the misalignment between the CMB and 2MRS dipoles.  
We discuss three possibilities.

\section{Explanations for the Discrepancy Between the CMB and 
2MRS Dipoles}

\subsection{Transverse Motion of the Milky Way}

The velocity estimate given in equation (\ref{vLGCMB1}) assumes that
the Milky Way has no transverse motion around the LG center of mass.
We investigate if such motion might explain the velocity discrepancy.

We begin by considering the component of $\vec{v}_{\LG-\CMB}$ towards
M31 (i.e., parallel to $\hat{n}_{{\rm M31}}$) 
since this component is independent of
the transverse velocity.  Equating the components of the velocities
(\ref{vLGCMB1}) and (\ref{vLGCMB2}) along this direction, we solve for
the bias factor to obtain $b=0.845\pm0.237$.  The central value is not
very likely since it is less than unity.  Nevertheless, we substitute
this estimate of the bias back into the two expressions for
$\vec{v}_{\LG-\CMB}$ to infer the transverse velocity of the Galaxy
around the LG center of mass:
\begin{equation}
\vec{v}_{\perp,\GC-\LG} =  (252.4\pm130.9, 61.7\pm155.8, -197.2\pm152.6) \kms.
\end{equation}
Including the reflex motion of M31, this estimate predicts the
following proper motion of M31,
\begin{equation}
\vec{v}_{\rm M31-MW} = (-429\pm223, -105\pm265, 335\pm259) \kms.
\end{equation}
If instead we use the estimate $b=1.056$ which we obtained in \S~3,
then
\begin{equation}
\vec{v}_{\rm M31-MW} = (-343\pm178, 98\pm212, 169\pm208) \kms.
\end{equation}
In either case, we see that we require a large transverse velocity of
M31 with respect to the Milky Way, whose most significant component is
a large velocity towards the Galactic Anticenter.

Loeb et al. (2005) constrained the proper motion of Andromeda to be $\sim
100\pm 20 \kms$ based on the measured proper motion of its satellite M33
and the requirement that M33 should not be tidally disrupted in the past.
Van der Marel \& Guhathakurta (2007) assumed that M31's satellites on
average follow Andromeda's motion relative to the Local Group; they
accordingly used the line-of-sight velocities of 17 satellites of Andromeda
and 5 galaxies at the outskirts of the Local Group, as well as the proper
motions of M33 and IC10, to infer $\vec{v}_{\rm M31-LG} = (97\pm35, -67\pm26,
80\pm32) \kms$. The transverse speed of Andromeda inferred by these studies
is well below the central value needed to explain the discrepancy between
the 2MRS and the CMB dipoles. Moreover, the $x$-component of the velocity
inferred by Van der Marel \& Guhathakurta (2007) is positive whereas the
CMB-LG discrepancy requires a large negative value.

As a side note, we use the Courteau \& van den Bergh (1999) estimate
of the Sun's motion relative to the LG to obtain the velocity of the
Galaxy with respect to the LG:
\begin{equation}
\vec{v}_{\GC-\LG} = (-57.8\pm26.5,59.6\pm18.3,-29.0\pm21.3) \kms.
\end{equation}
This gives a speed along the GC-LG direction of $85.9\pm20.9 \kms$,
which is statistically consistent with our previous more accurate
estimate of $66.6\pm6.7 \kms$.  For the transverse speed, the estimate
of Courteau \& van den Bergh (1999) gives $18.6\pm32.4 \kms$, which is
again much smaller than the velocity discrepancy we seek to explain.

In the next two subsections we assume that the Milky Way has
negligible velocity transverse to the LG center of mass and consider
whether incompleteness in the 2MRS might explain the misalignment
between the CMB and 2MRS dipoles.

\subsection{Structure Beyond the Maximum Distance of 2MRS}

The 2MRS sample of Erdo{\u g}du et al. (2006) extends out to a velocity of
$20,000 \kms$.  Figure 6 of their paper shows that the flux-weighted dipole
in the CMB frame receives most of its contribution from inside about $4,000
\kms$, which is much shorter than the limiting distance of the survey.
This suggests that any contribution from beyond the survey volume is likely
to be quite small.

To verify this, we considered two logarithmically spaced velocity bins in
the 2MRS sample: Bin I, $5,000$--$10,000\kms$, and Bin 2,
$10,000$--$20,000\kms$.  From the data given in Table 1 of Erdo{\u g}du
et al. (2006), we computed the mean square contribution to the quantity
$(b/f(\Omega_m)) v_{\LG-\CMB}$ from each of the two bins.  We obtained
$(162\kms)^2$ and $(64\kms)^2$ from Bins 1 and 2, respectively.  The
numerical estimates are consistent with a scale-invariant Cold Dark Matter
(LCDM) power spectrum in the linear regime, for which the mean square
velocity should vary roughly as the inverse square of the distance (Peacock
1998).  We then estimate the root mean square contribution from the rest of
the universe beyond $20,000\kms$ to be $(b/f(\Omega_m))
v_{\LG-\CMB}\sim 80\kms$.  For $f(\Omega_m)/b \approx 0.4$ (Erdo{\u g}du et
al. 2006), this gives $v_{\LG-\CMB}(>20,000\kms)\sim 30\kms$, which is
very much smaller than the $\sim 200\kms$ we need to eliminate the
misalignment between the CMB and 2MRS dipoles (eq. \ref{vLGCMB3}).

Peacock (1992) analysed the expected convergence of the dipole velocity
with distance based on the large-scale power spectrum. He finds that the
misalignment angle between the true CMB dipole and the dipole measured
within a finite survey volume is expected to be negligible beyond a
distance of $20,000\kms$ ($\sim 280$Mpc). Hence, distant structure beyond
the limit of 2MRS is very unlikely to be the source of the inferred
discrepancy between the CMB and 2MRS dipoles.

\subsection{Nearby Structure in the ZOA}

Finally, we consider the possibility that nearby galaxies inside the
ZoA may be responsible for the discrepancy between the CMB and 2MRS
dipoles.  Erdo{\u g}du et al. (2006) state that the ZoA for their
survey corresponded to the area $|b|<5^\circ$ for $|l|>30^\circ$ and
$|b|<10^\circ$ for $|l|<30^\circ$.  Given this information plus their
estimate of the contribution to $\vec{v}_{\LG-\CMB}$ from the region
of the sky covered by their survey, we estimate the {\it
root-mean-square} contribution of the ZoA to each of the components of
the LG velocity to be,
\begin{eqnarray}
\sigma_{\ZoA,x} &=& 168.7/b \kms, \label{ZoAx} \\
\sigma_{\ZoA,y} &=& 150.2/b \kms, \label{ZoAy} \\ 
\sigma_{\ZoA,z} &=& 15.5/b \kms. \label{ZoAz}
\end{eqnarray}
As expected, the contribution in the $z$ direction is small.

Multiplying equation (\ref{vLGCMB1}) by $b$ and subtracting from
equation (\ref{vLGCMB2}), we obtain the following estimates for the
contribution of the ZoA to the velocity of the LG,
\begin{eqnarray}
bv_{\ZoA,x}\!\!\!&=&\!\!\!214.8-1.8b;
\ \sigma=\left(110.6^2+3.5^2b^2\right)^{1/2}, \\
bv_{\ZoA,y}\!\!\!&=&\!\!\!506.1-537.2b; 
\ \sigma=\left(131.1^2+14.1^2b^2\right)^{1/2}, \\
bv_{\ZoA,z}\!\!\!&=&\!\!\!-414.4+293.2 b;  
\ \sigma=\left(128.9^2+4.0^2b^2\right)^{1/2},
\end{eqnarray}
where the {\it root-mean-square} uncertainty in each expression is
given by the $\sigma$ value on the right, and all quantities are in
$\kms$.  These three equations can be used to derive an expression for
the likelihood of the three velocity components.  Before writing this
likelihood function we note that we have calculated in equations
(\ref{ZoAx})--(\ref{ZoAz}) the {\it root-mean-square} expectation
values of $bv_{\ZoA,x}$, $bv_{\ZoA,y}$ and $bv_{\ZoA,z}$, which supply
us with the prior distributions of these three velocities.  In
addition, we have a fourth unknown quantity, namely the bias factor
$b$, for which we adopt a flat prior.

We thus obtain the following likelihood function for the four
unknowns,
\begin{eqnarray}
& &P(bv_{\ZoA,x},b v_{\ZoA,y},bv_{\ZoA,z},b)
\nonumber \\ 
& &\propto \exp\left[-{(b v_{\ZoA,x})^2\over 2(168.7)^2}-
{(bv_{\ZoA,x}-214.8+1.8b)^2\over2(110.6^2+3.5^2b^2)}\right] \times
\nonumber \\ &~& \exp\left[-{(b v_{\ZoA,y})^2\over 2(150.2)^2}-
{(bv_{\ZoA,y}-506.1+537.2b)^2\over2(131.1^2+14.1^2b^2)}\right] \times
\nonumber \\ &~& \exp\left[-{(b v_{\ZoA,z})^2\over 2(15.5)^2}-
{(bv_{\ZoA,z}+414.4-293.2b)^2\over2(128.9^2+4.0^2b^2)}\right] 
. 
\label{likelihood}
\end{eqnarray}
By marginalizing this likelihood over any three of the four unknown
quantities, we obtain the probability distribution for the fourth.

The results are shown by the solid lines in Fig. 1.  We find that the
bias factor has a fairly broad distribution with a $1\sigma$ range
from $\sim0.85$ to $\sim1.4$.  Since it is most unlikely that the
galaxies detected by 2MASS would have a bias less than unity, we have
repeated the calculations with a prior for $b$ truncated below unity.
The corresponding results are shown by the dotted lines.

\begin{figure}
\begin{center}
\includegraphics[width=8.5cm]{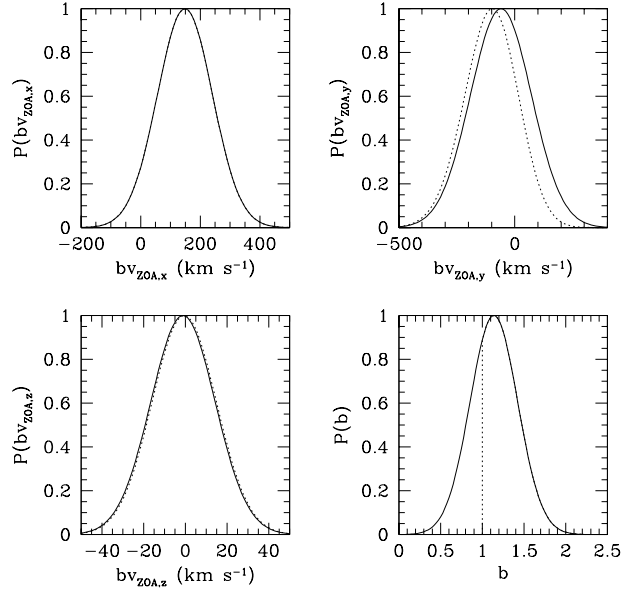}
\caption{Marginalized probability distributions of $bv_{\ZoA,x}$,
$bv_{\ZoA,y}$, $bv_{\ZoA,z}$ and $b$.  The distributions have been
normalized such that their maxima are equal to unity.  The solid lines
correspond to a flat prior for $b$ extending from 0 to infinity, while
the dotted lines correspond to a prior in which $b$ is restricted to
be $\geq1$}
\end{center}
\end{figure}

Figure 1 indicates that $bv_{\ZoA,x}$ has a $95\%$ probability of
being positive (a strong $2\sigma$ result), while $bv_{\ZoA,y}$ has a
$68\%$ probability of being negative (a weaker $1\sigma$ result).  The
most likely values of these two components are $bv_{\ZoA,x}\sim 150
~\kms$ and $bv_{\ZoA,y}\sim -60 ~\kms$, though each has a broad
probability distribution.  When we restrict the bias factor to $b \geq
1$ (dotted lines in Fig. 1), the corresponding numerical results are
$95\%$, $82\%$, $150 ~\kms$ and $-100 ~\kms$, respectively.  The
velocity component $bv_{\ZoA,z}$ is consistent with zero, as expected.

In contrast to our analysis in \S\S~4.1
and 4.2, where the particular explanations considered there 
were easily ruled out,
now we see that acceleration from galaxies in the ZoA may well explain
the misalignment between the CMB and 2MRS dipoles.  The magnitude of the
velocity discrepancy is consistent with the expected contribution from
the ZoA (described by our estimates of $\sigma_{\ZoA,x-z}$).
Moreover, the additional acceleration from the ZoA is expected to be
in the $x$--$y$ plane, and most likely in the $x$ direction, i.e.,
towards the Galactic Center where obscuration is maximum.  This is
exactly the sense of the velocity discrepancy.  
Given this encouraging agreement, we predict that a
survey of the ZoA would find additional structure in the nearby
universe, especially behind the Galactic Center region.

\begin{figure}
\begin{center}
\includegraphics[width=8.5cm]{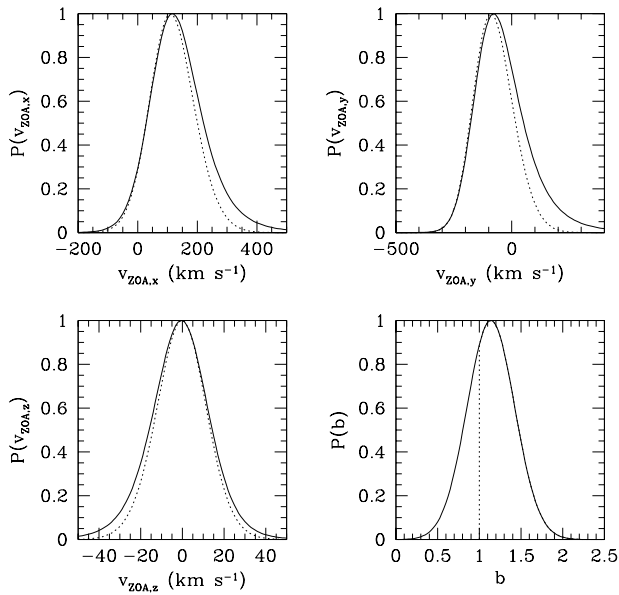}
\caption{Marginalized probability distributions of $v_{\ZoA,x}$,
$v_{\ZoA,y}$, $v_{\ZoA,z}$ and $b$.  The distributions have been
normalized such that their maxima are equal to unity.  The solid and
dotted lines are as in Fig. 1.  The distribution of $b$ is identical
to the one shown in Fig. 1.}
\end{center}
\end{figure}

Figure 1 corresponds to the probability distributions of the
components of the {\it bias-multiplied} velocity $b\vec{v}_{\ZoA}$
since these quantities are most directly related to the 2MRS survey.
For completeness we show in Fig. 2 the distributions of the velocity
components themselves.  These were calculated in the same way, except
that we considered the likelihood function $P(v_{\ZoA,x}, v_{\ZoA,y},
v_{\ZoA,z}, b)$.  This quantity is almost the same as the likelihood
given in equation (\ref{likelihood}) except that it differs by the
following Jacobian,
\begin{equation}
J \equiv {\partial(bv_{\ZoA,x}, bv_{\ZoA,y}, bv_{\ZoA,z}, b) \over
\partial(v_{\ZoA,x}, v_{\ZoA,y}, v_{\ZoA,z}, b)} = b^3.
\end{equation}

Figure 2 is generally consistent with the results in Fig. 1.  For
completeness we note the following numerical results: The probability
of $v_{\ZoA,x}$ being positive is $95\%$ and the most likely value
of this velocity component is $120 ~\kms$ (solid line) and $110 ~\kms$
(dotted line).  The probability of $v_{\ZoA,y}$ being negative is
$68\%/82\%$ (solid/dotted line) and the most likely value is $-80/-90
~\kms$ (solid/dotted line).

The 2MRS results that we have used from Erdo{\u g}du et al. (2006)
correspond to their ``second method'' of treating the ZoA, in which
they fill the ZoA with structure consistent with that found in
neighboring latitude strips.  We do not know how effective this method
is at predicting the missing information.  If it were perfect, there
should be no discrepancy between 2MRS and the direction of the CMB
dipole.  In our analysis, we have assumed that 2MRS has no information
at all inside the ZoA.  To be consistent with this assumption, we
should ideally use the results corresponding to Erdo{\u g}du et al.'s
``first method,'' in which the authors simply fill the ZoA with random
galaxies.  Unfortunately, their paper does not give a table of results
corresponding to this method.

\section{Discussion}

We have obtained new constraints on the dynamics of the Local Group by
comparing the CMB and the 2MRS dipoles. The analysis is limited by the
lack of surveyed galaxies behind the Zone of Avoidance (ZoA).  In
order to match the CMB and 2MRS dipoles, we have inferred excess
motion (that is not acounted for by 2MRS) along the direction of the
Galactic center (\S 4.2). This happens to be the most natural
direction for hiding mass concentrations because it is associated with
enhanced confusion and dust obscuration by the Galaxy.

The implications of our findings have a simple interpretation.  To acquire
an excess peculiar velocity $\Delta v$ towards the ZoA over the age of the
Universe, $\Delta t=1.4\times 10^{10}~{\rm yr}$, requires an average
acceleration of order $g \sim \Delta v/\Delta t$.  Assuming that this
acceleration is induced gravitationally (i.e. $g\sim GM/d^2$) by a hidden
object of mass $M$ at a characteristic distance $d$, we get the simple
scaling relation for the required mass: $M_{12} \sim 1.7 v_7 d_0^2$,
where $M_{12}=(M/10^{12}M_\odot)$, $v_7=(\Delta v/100~{\rm km~s^{-1}})$ and
$d_0=(d/1~{\rm Mpc})$.  According to Fig. 2, the most likely value of the
excess velocity towards the GC is $\Delta v\equiv v_{\rm ZOA,x}\sim
120~{\rm km~s^{-1}}$, which requires a hidden galaxy comparable in mass to
Andromeda ($M\sim 2\times 10^{12}M_\odot$) at a distance of $\sim 1$ Mpc,
or a hidden galaxy cluster comparable in mass to the Coma cluster ($M\sim
10^{15}M_\odot$) at a distance of $\sim 20$ Mpc. At these distances, the
inner 10 kpc diameter of a hidden galaxy would occupy an angle of $\sim
0.6^\circ$, and the inner 1Mpc diameter of a hidden cluster would occupy
$\sim 2.5^\circ$.  An extended supercluster might not be fully hidden
behind the ZoA.  As argued in \S 4.2, it is very unlikely that structure
beyond the maximum distance of 2MRS of $\sim 280$ Mpc ($\sim 20,000
\kms$) accounts for the discrepancy. The LCDM power spectrum would
typically account for a velocity offset that is an order of magnitude
smaller than the central value we infer.

The above possible objects are already constrained by existing data. In
particular, a new galaxy cluster must have escaped detection by dedicated
ZoA searches in the X-ray band (Ebeling et al 2002, 2005; Kocevski 2005),
optical galaxy searches (Roman et al 1998; Wakamatsu et al. 2005; Hasegawa
et al. 2000) and 21cm surveys (Meyer et al 2004; Henning et al.2005;
Kraan-Korteweg et al. 2007). Similarly, a nearby galaxy must have escaped
detection by 2MRS as well as existing 21cm surveys.  We are not in a 
position to evaluate how likely this is.

We note that the excess mass we predict behind the GC will have an 
associated infrared flux which is independent of the
attractor's mass and is linearly proprtional to $\Delta v$, since it scales
as $\propto M/d^2$ for a fixed mass-to-light ratio. Erdo{\u g}du et
al. (2006) infer a net luminosity density of $(7.67\pm 1.02)\times 10^8 h
L_\odot~{\rm Mpc^{-3}}$ for the 2MRS galaxies. When compared to the average
matter density of the Universe for $\Omega_m=0.24$ and $h=0.7$, this
results in a predicted excess radiation flux of $\sim 2.4\times 10^{9} v_7
L_\odot{\rm ~Mpc^{-2}}$ behind the GC.

Unfortunately, the current error budget is too large to provide a
useful constraint on the transverse velocity of the Milky Way relative
to the Andromeda galaxy (\S 4.1).  A future survey of the ZoA might
allow to determine this transverse velocity by requiring a match
between the peculiar velocity inferred for the local group and the CMB
dipole.  The bias parameter of the surveyed galaxies, $b$, could be
determined by requiring that the inferred radial velocity of Andromeda
relative to the Milky-Way will match its observed value. It would then
be possible set a lower limit on the local group mass so that the two
galaxies will be gravitationally bound.

Measuring the relative transverse velocity of the Milky-Way and Andromeda
is of great interest, since it would affect the future trajectory of the
two galaxies (Cox \& Loeb 2007) and will determine whether the LG is likely
to be gravitationally bound (Binney \& Tremaine 1986). Current methods for
inferring the transverse speed (Loeb et al. 2005; van der Marel \&
Guhathakurta 2007) are indirect and highly uncertain.  A future ZoA survey
for infrared or 21cm galaxies (Henning et al. 2005) would provide a new
elegant path for constraining this unknown velocity component, which is
difficult to measure otherwise.

\bigskip

{\bf Acknowledgments} The authors thank John Huchra and Mark Reid for
useful discussions. We also thank an anonymous referee for a number of
very valuable comments and for pointing out the existing constraints
on nearby objects behind the ZoA.

\newcommand{\noopsort}[1]{}

\label{lastpage}
\end{document}